\documentclass[prl,aps,reprint,twocolumn,showpacs,superscriptaddress]{revtex4-2}
\usepackage{physics}
\usepackage[english]{babel}
\usepackage[utf8]{inputenc}
\usepackage[pdftex]{graphicx}
\usepackage{dsfont}
\usepackage{bm}
\usepackage[normalem]{ulem}
\usepackage{hyperref}
\usepackage{epstopdf}
\usepackage{xcolor}
\usepackage[sort&compress]{natbib}
\usepackage{subfigure}
\usepackage{colordvi}
\usepackage{epsfig}
\usepackage{float}
\usepackage{dcolumn}
\usepackage{ifpdf}
\usepackage{graphicx,color,bm}
\usepackage[english]{babel}
\usepackage{amsmath,amssymb}
\usepackage{stmaryrd,dsfont,txfonts}
\usepackage{subfigure}
\usepackage{hyperref}
\hypersetup{
	colorlinks,%
	citecolor=blue,%
	linkcolor=blue,%
	urlcolor=blue
}
\providecommand{\gerson}[1]{\textcolor{black}{#1}}
\providecommand{\fu}[1]{\textcolor{black}{#1}}

\providecommand{\fufu}[1]{\textcolor{black}{#1}}
\providecommand{\fufufu}[1]{\textcolor{black}{#1}}

\begin{document}
	\title{An Unusual  Dresselhaus Spin-Orbit Contribution of  Even Order in Momentum}
	\author{Hao Yang}
	\affiliation{Department of Physics, Qufu Normal University,  Qufu 273165, Shandong, China}
                  \author{Wei Wang}
                   \affiliation{Department of Physics, Jining University, 273155 Qufu, Shandong, China}
      \author{Gerson J. Ferreira }
      \affiliation{Instituto de F\'{\i}sica, Universidade Federal de Uberl\^andia, Uberl\^andia, Minas Gerais 38400-902, Brazil}
         \author{Ning Hao}
         \affiliation{Anhui Province Key Laboratory of Low-Energy Quantum Materials and Devices, High Magnetic Field Laboratory, HFIPS, Chinese Academy of Sciences, Hefei, Anhui 230031, China}
           \author{Ping Zhang}
\affiliation{Department of Physics, Qufu Normal University,  Qufu 273165, Shandong, China}
\affiliation{Beijing Computational Science Research Center, Beijing 100084, China}
	\author{Jiyong Fu}
	\thanks{yongjf@qfnu.edu.cn}
	\affiliation{Department of Physics, Qufu Normal University, Qufu 273165, Shandong, China}
        \begin{abstract}
         The spin-orbit  (SO) coupling is conventionally known to manifest  as \emph{odd} functions of  momentum.
         Here, \fu{through both model calculations and symmetry analysis along with \gerson{the method of invariants},} we reveal that, in ordinary
         semiconductor  \fufu{heterostructures}, a \emph{quadratic} Dresselhaus SO term---inheriting from its bulk crystal form---emerges via the interband  effect, while  complying   with time-reversal \gerson{and spatial symmetries}.  Furthermore, we observe that this  unusual SO term  gives rise to a range of striking quantum phenomena, including  hybridized swirling texture, anisotropic energy dispersion,  avoided band crossing,   {longitudinal} \emph{Zitterbewegung},  and opposite  spin evolution between different bands in quantum dynamics.   These  stand in stark contrast to those associated with  the usual \emph{linear} SO \gerson{terms}.
         Our findings uncover a previously overlooked route for exploiting interband effects and open new avenues for spintronic functionalities that leverage
         unusual  SO terms of \emph{even} orders  in momentum.
          \end{abstract}
\maketitle

\paragraph*{Introduction.---}
The spin-orbit (SO) interaction, which facilitates coherent spin manipulation, is a key ingredient in spintronic devices~\cite{awschalom:2002,zutic:2004,RevModPhys.96.021003, fabian:2007,Fabian:2009,bindel:2016, rodriguez2024magnetic,nitta2023spin}.
In semiconductor nanostructures, the SO effects usually have two dominant contributions:  the Rashba~\cite{rashba:1984} 
and Dresselhaus~\cite{dresselhaus:1955}   terms, arising from the structural and crystal  inversion asymmetries, respectively.
Constrained by time-reversal symmetry (TRS), the SO coupling  is conventionally expected to appear as \emph{odd} functions of momentum.
The Rashba term is \emph{linear}, and  can be tuned with the doping profile~\cite{koralek:2009} as well as \emph{in situ} using gate voltages~\cite{engels:1997,nitta:1997}.
The Dresselhaus term inherits  its  bulk three-dimensional (3D) source but with a reduced  two-dimensional (2D) form~\cite{dettwiler:2017,fu:2016,fu:2015}, which  can be  decomposed into both \emph{linear} and  \emph{cubic} terms of odd orders in momentum~\cite{studer:2010,dettwiler:2017,fu:2016,fu:2015,yang2025emergence}.

\begin{figure}[h]
	\includegraphics[width=7.6cm]{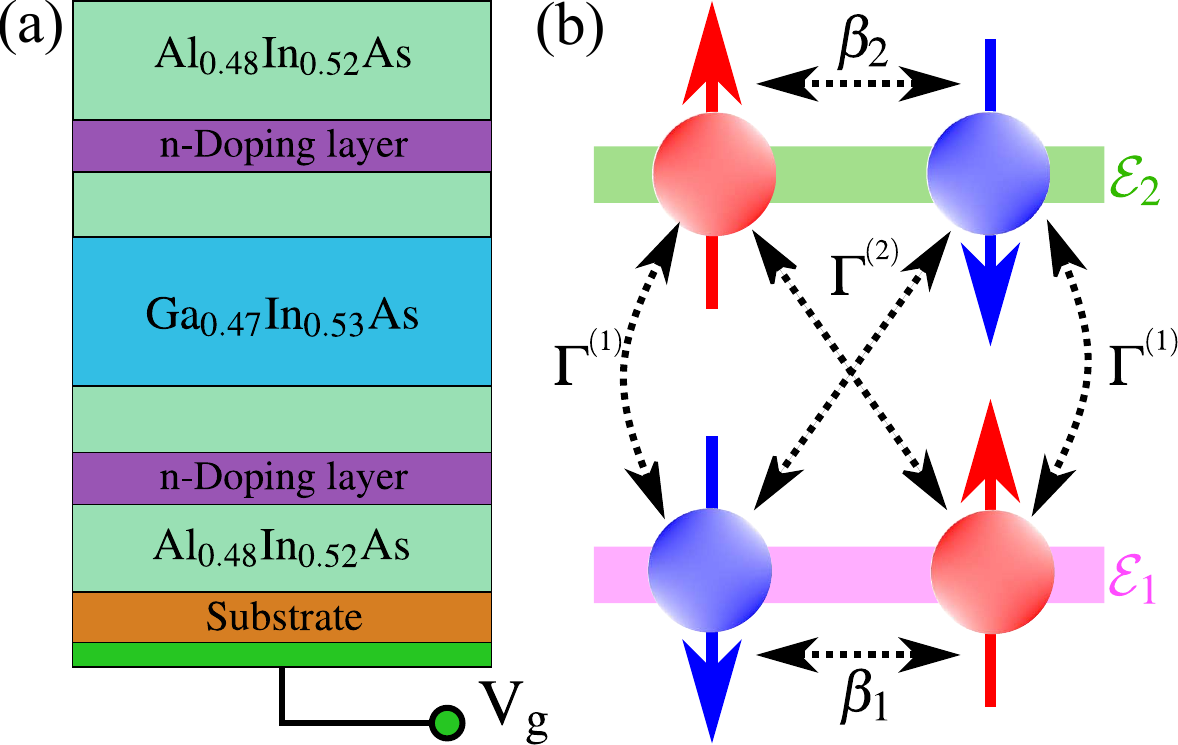}
	\centering
	\caption{(Color online) (a) Growth profile of an \emph{n}-doped
		${\rm Al_{0.48}In_{0.52}As/Ga_{0.47}In_{0.53}As}$ \fufu{quantum well} subject to gate voltage $V_{\rm g}$, and the potential profile
	  for 2DEGs  with two-band  energies $\varepsilon_1$ and $\varepsilon_2$.
          (b) Illustration of intra- and interband Dresselhaus SO couplings with spin and band degrees of freedom. Here $\beta_{\nu}$ $(\nu=1,2)$ denotes the intraband SO connecting distinct spins within the same $\nu$-th band, and $\Gamma^{(1)}$ ($\Gamma^{(2)}$) represents the linear (quadratic)  interband SO
          between spins of opposite (same) orientations. The red and blue arrows stand for the spin-up and spin-down states, respectively. }
	\label{fig1}
\end{figure}

\begin{figure}[htb!]
	\includegraphics[width=8.0cm]{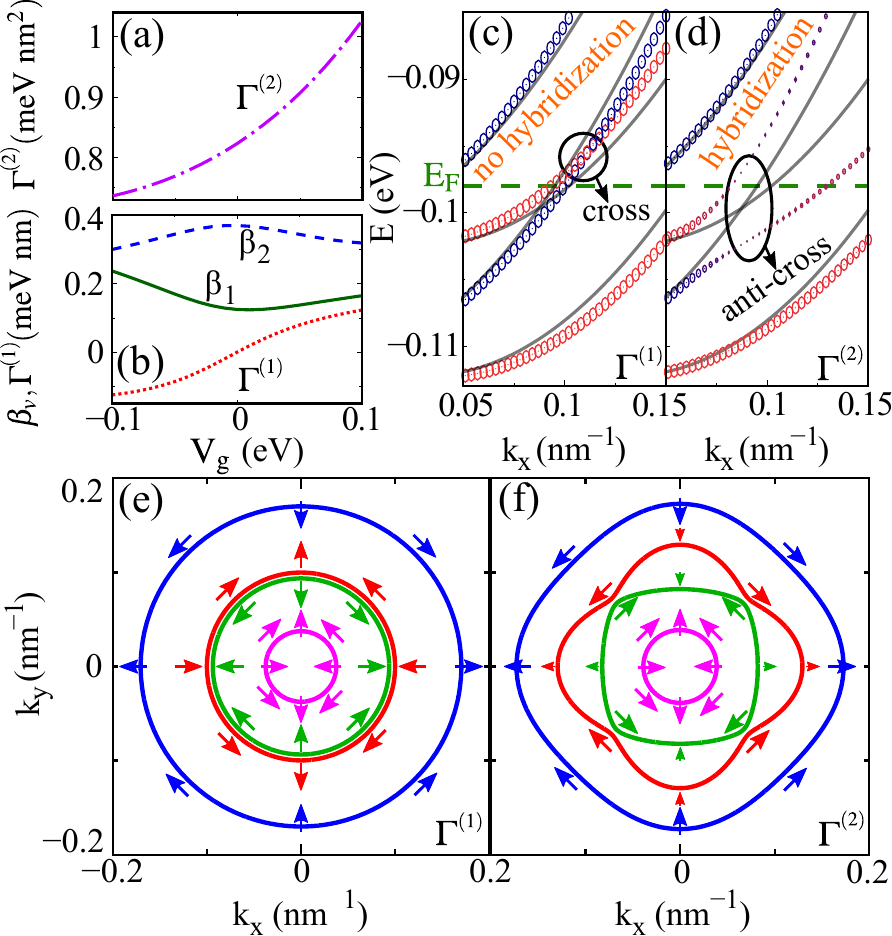}
	\centering
	\caption{(Color online) (a) The  quadratic (interband $\Gamma^{(2)}$)  and (b) the linear (intraband $\beta_\nu$ and interband $\Gamma^{(1)}$)  Dresselhaus SO coefficients versus $V_{g}$. \gerson{(c), (d)} Spin-resolved two-band energy dispersion  (scaled by a factor of 100 for visibility) along the $k_x$ direction, with interband coupling $\Gamma^{(1)}$ \gerson{(c)} and $\Gamma^{(2)}$ \gerson{(d)}.  The size of markers scales with the degree of spin
	  polarization and the color represents spin orientations up (red) and down (blue).
          The gray (dashed) lines indicate the uncoupled ($ \Gamma^{(1)}=\Gamma^{(2)}=0$) bands.
          \gerson{(e), (f)} Constant-energy contours and spin textures at \fufu{${E_{\rm F}=-0.098}$} eV
          [see horizontal (green) dashed line in (c) and (d)], for our system with $\Gamma^{(1)}$ (e) and $\Gamma^{(2)}$ (f).
          }
\label{fig2}
\end{figure}

\begin{figure}[htb!]
	\includegraphics[width=8.0cm]{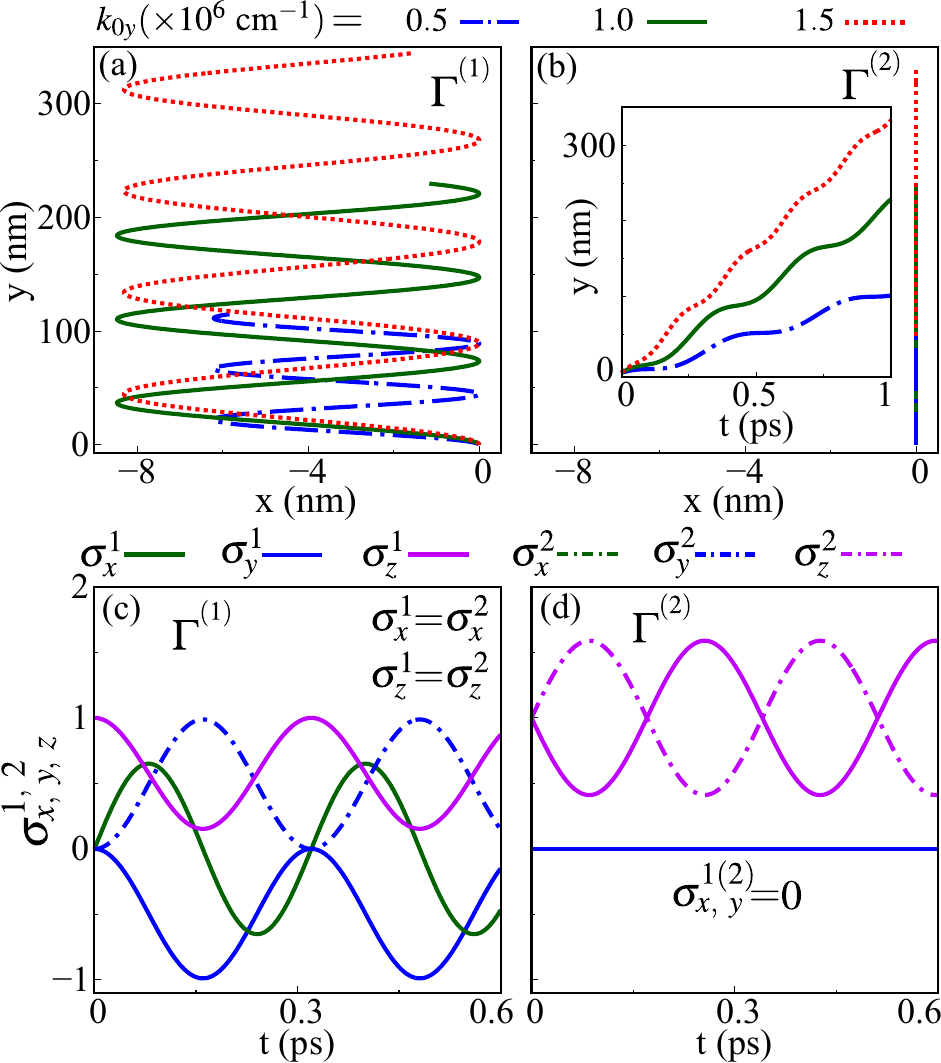}
	\centering
	\caption{(Color online) (a), (b) \textit{Zitterbewegung} induced by the linear ($\Gamma^{(1)}$) (a) and quadratic ($\Gamma^{(2)}$) (b) interband  SO
          terms, for the \emph{initial} (group) velocity $\mathbf{v}_{\rm g}=\hbar k_{0y}/m^*\hat{\mathbf{y}}$  (along the $y$  direction) with  $k_{0y}=\{0.5, 1.0, 1.5\}\times 10^{6}$ cm$^{-1}$. In (b), the $x$-direction motion vanishes for all $k_{0y}$;  the inset shows the $y$-direction
          trajectory over time. (c), (d) Time evolution of spin components $\sigma_{x,y,z}^\nu$ ($\nu=1,2$) for the two bands in the presence of  $\Gamma^{(1)}$ (c) and $\Gamma^{(2)}$ (d), at $k_{0y}=1.0\times 10^6$ cm$^{-1}$. }
	\label{fig3}
\end{figure}

\begin{figure}[htb!]
	\includegraphics[width=8.0cm]{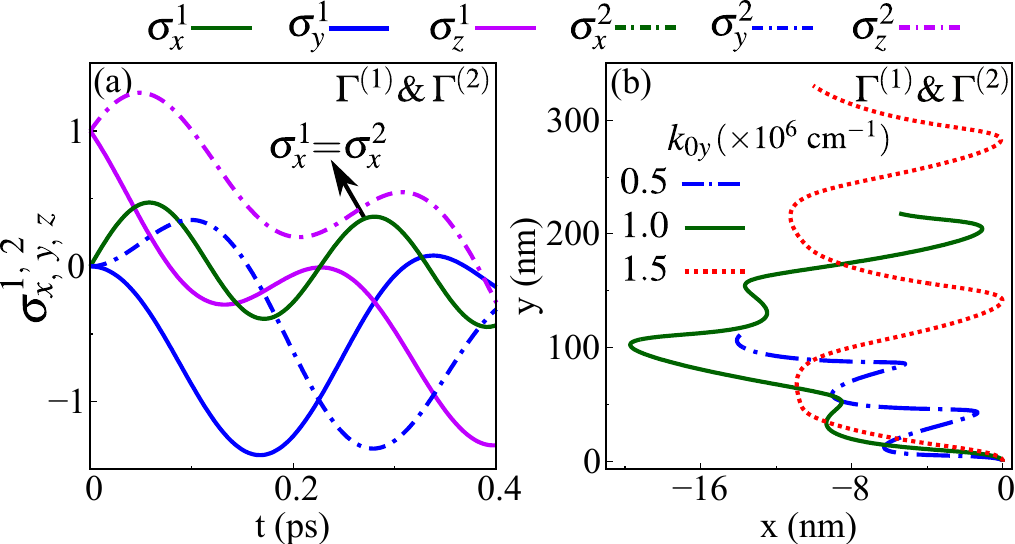}
	\centering
	\caption{(Color online)  (a), (b) Time evolution of the two-band spin components $\sigma_{x,y,z}^\nu$ (a) and \emph{Zitterbewegung} (b),
          incorporating both linear ($\Gamma^{(1)}$) and quadratic ($\Gamma^{(2)}$) SO contributions. In (a), $k_{0y}=0.05\times 10^6$ cm$^{-1}$;
          in (b),  $k_{0y}=\{0.5, 1.0, 1.5\}\times 10^{6}$ cm$^{-1}$.} 
	\label{fig4}
\end{figure}

Here,  for two-dimensional electron gases (2DEGs) confined in ordinary semiconductor  heterostructures [Fig.~\ref{fig1}(a)],
we demonstrate that the interband coupling enables the emergence of a \emph{quadratic} [$\Gamma^{(2)}$; Fig.~\ref{fig1}(b)] Dresselhaus SO term, while remaining fully consistent with TRS. \fu{ This unusual SO term is supported by both our model calculations and symmetry analysis along with \gerson{the method of invariants \cite{bir1974}}.} Furthermore, we observe  that the  unusual  SO term induces diverse spin-related phenomena,  including  hybridized swirling  texture, anisotropic energy dispersion,  avoided band crossing [Fig.~\ref{fig2}],   {longitudinal} \emph{Zitterbewegung},  and opposite  spin evolution between different bands in quantum dynamics  [Figs.~\ref{fig3} and \ref{fig4}].  All these stand in stark contrast to those associated with  the usual \emph{linear} SO term.
Our work uncovers a previously overlooked facet of SO  physics in nanostructures and points toward novel strategies for
 spintronic applications that leverage interband effect and  unusual  SO terms of \emph{even} orders.

\paragraph*{Model Hamiltonian.---} {We start from} the bulk Dresselhaus cubic SO coupling in zincblende crystals~\cite{dresselhaus:1955},
\begin{equation}
\mathcal{H}_{\mathcal{D}}^{\text {bulk }}=(\gamma/\hbar)[\sigma_x p_x(p_y^2-p_z^2)+ \rm{c.p.} ],
\label{bulk_Hamiltonian}
\end{equation}
with $p_{x, y, z}$ being the electron momentum components along the $x\parallel[100]$, $y \parallel[010]$ and $z\parallel[001]$ directions, $\sigma_{x, y, z}$ refering to the Pauli matrices, and $\gamma$ as the bulk Dresselhaus parameter.

We consider 2DEGs  confined in semiconductor \fufu{quantum wells} grown along the $z$ direction. Hence, we have  the overall 3D Dresselhaus SO Hamiltonian,
\begin{equation}
H=H_{\rm 0} +   \mathcal{H}_{\mathcal{D}(1)}^{3 \mathrm{D}}+\mathcal{H}_{\mathcal{D}(2)}^{3 \mathrm{D}}+\mathcal{H}_{\mathcal{D}(3)}^{3 \mathrm{D}},
\label{full_Hamiltonian}
\end{equation}
in which $H_{\rm 0}=\hbar^2(k_{x}^2+ k_{y}^2)/{2m^*} +k_z(\hbar^2/2{m^*)k_z} + V_{sc}(z)$, $m^* \equiv m^*(z)$ is the $z$-dependent effective electron mass and $k_z=-i \partial_z$. Here $V_{sc}$ denotes the self-consistent potential felt by the confined 2DEGs. It comprises the structural part $V_w$,  the electronic Hartree potential $V_e$, the doping potential $V_d$ and an external gate $V_g$.  The other terms in $H$, namelly $\mathcal{H}_{\mathcal{D}(i)}^{3 \mathrm{D}}$, denote the $i$-th order Dresselhaus SO contribution, with  $\mathcal{H}_{\mathcal{D}(1)}^{3 \mathrm{D}}=k_z\gamma(z) k_z(\sigma_y k_y-\sigma_x k_x)$, $\mathcal{H}_{\mathcal{D}(2)}^{3 \mathrm{D}}=(1/2)(\gamma(z) k_z+k_z\gamma(z)) \sigma_z(k_x^2-k_y^2)$, and $\mathcal{H}_{\mathcal{D}(3)}^{3 \mathrm{D}}=\gamma(z) k_x k_y(\sigma_x k_y-\sigma_y k_x)$. Note that here
we have performed the usual  \emph{symmetrization} procedure~\cite{bastard:1989} since $\gamma(z)$ and $k_z$ do not commute. 

We now derive an effective 2D Hamiltonian by projecting $H$ [Eq.~\eqref{full_Hamiltonian}] onto the {\it two} lowest spin-degenerate eigensolutions of $H_{\rm 0}$: $\langle \mathbf{r}| \mathbf{k},\nu,\sigma\rangle=e^{i\mathbf{k}\cdot\mathbf{r}}\varphi_\nu(z)|\sigma_z\rangle$, \gerson{$\nu=\{1,2\}$, $\sigma_z = \{\uparrow, \downarrow\}$},
with energies \fufufu{$\varepsilon_{\nu,k} = \varepsilon_{\nu} + \hbar^2k^2/2m_{\rm 2D}^*$}, where $\mathbf{k} = (k_x, k_y)$ is the in-plane wave vector, $\varepsilon_\nu$ is the $\nu$th (self-consistent) energy level, \fufufu{and $m_{\rm 2D}^*$ denotes the effective 2D electron mass}~\cite{footnote-2dmass}. We find the effective $4\times4$ 2D Hamiltonian~\cite{footnote-second-order},
\begin{equation}
\fufufu{\mathcal{H} = \left(\dfrac{\hbar^2k^2}{2m_{2D}^*}+ \varepsilon_+\right)\mathds{1}\otimes\mathds{1} }-\varepsilon_-\tau_z\otimes\mathds{1}
+ \mathcal{H}_{\rm D},
\label{fullH}
\end{equation} 
in which $\mathds{1}$ the 2 $\times$ 2 identity matrix, $\varepsilon_{\pm}=\left(\varepsilon_2\pm\varepsilon_1\right)/2$, $\tau_j$ ($j=x,y,z$) the Pauli matrices in the subband subspace, and $\mathcal{H}_{\rm D} $ the two-band Dresselhaus SO Hamiltonian,
\begin{eqnarray}
	\mathcal{H}_{\rm D} &=& \frac{g^*\mu_B}{2}\left[\right. \mathds{1}\otimes\bm{\sigma}\cdot\mathbf{B}^{\rm D}_+(\mathbf{k})
	-\tau_z\otimes\bm{\sigma}\cdot\mathbf{B}^{\rm D}_-(\mathbf{k}) \nonumber \\
     &+&  \tau_{x}\otimes\bm{\sigma}\cdot\mathbf{B}^{\rm D(1)}_{12}(\mathbf{k})-\tau_{y}\otimes\bm{\sigma}\cdot\mathbf{B}^{\rm D(2)}_{12}(\mathbf{k})\left.\right],
	\label{H_SO}
\end{eqnarray}
where $g^*$ is the effective g-factor, $\mu_B$ is the Bohr magneton, $\mathbf{B}_{\pm}^{\rm D}=(\mathbf{B}^{\rm D}_{2}\pm\mathbf{B}^{\rm D}_{1})/2$, and
$\mathbf{B}^{\rm D}_{\nu}$ is the \emph{intraband} Dresselhaus field,
\begin{eqnarray}
\mathbf{B}^{\rm D}_{\nu}=\dfrac{2k\beta_\nu}{g^*\mu_B}
\left(\sin\theta  \hat{\mathbf{y}}-\cos\theta \hat{\mathbf{x}} \right),
\label{intrafield}
\end{eqnarray}
with $\tan\theta=k_{y}/k_{x}$ and the {intraband} Dresselhaus strength $\beta_v=\langle v| \partial_z\gamma(z) \partial_z|v\rangle$ [Fig.~\ref{fig1}(b)]. 
 The linear ($\mathbf{B}^{\rm D(1)}_{12}$) and quadratic ($\mathbf{B}^{\rm D(2)}_{12}$) \emph{interband} Dresselhaus fields read as
\begin{equation}
\begin{split}
\mathbf{B}^{\rm D(1)}_{12} &= \dfrac{2k\Gamma^{(1)}}{g^*\mu_B}
\left(\sin\theta  \hat{\mathbf{y}}-\cos\theta \hat{\mathbf{x}}\right),
\\
\mathbf{B}^{\rm D(2)}_{12} &= \dfrac{2k^2\Gamma^{(2)}}{g^*\mu_B} \cos2\theta\hat{\mathbf{z}},
\end{split}
\label{interfield}
\end{equation}
with $\Gamma^{(1)}=\gamma\langle 1| \partial_z\gamma(z)\partial_z|2\rangle$ [$\Gamma^{(2)}=-(1/2)\langle 1| \gamma(z)\partial_z+\partial_z\gamma(z)|2\rangle$] as the linear [quadratic]  SO coefficients [Fig.~\ref{fig1}(b)].

\paragraph*{Symmetry analysis and invariants method.---}To corroborate our model, we further perform symmetry analysis~\cite{varjas2018qsymm} along with theory of invariants.  Zincblende crystals belong to the space group ${\rm F}\bar{4}3{\rm m}$~\cite{dresselhaus2007group}, which comprises the ${\rm T_{d}}$ point group and \gerson{nonsymmorphic} space translations. For \fufu{heterostructures} grown along the $z$ direction, the three-fold rotation ${\rm C_{3}^{[111]}}$ inherent to ${\rm T_{d}}$ is broken, and then
the point group is reduced to either ${\rm D_{2d}}$
or ${\rm C_{2v}}$, depending on whether the system is symmetric or asymmetric.
\fufu{For a general case with asymmetric configuration}, the relevant symmetry reduces to the point group
${\rm C_{2v}}$ along with time-reversal symmetry $\mathcal{T}$.
Within the method  of invariants~\cite{bir1974},  the effective 2D Hamiltonian  $\mathcal{H}(\mathbf{k})$ must satisfy $\mathcal{H}(D^{k}(S)\mathbf{k})=D^{\psi}(S)\mathcal{H}(\mathbf{k}){D^{\psi}}^{-1}(S)$,
for every symmetry operation $S \in {\rm C_{2v}\oplus\mathcal{T}}$, where $D^\psi$  stands for  the representation of basis functions and $D^k$ the representation acting on $\mathbf{k}$.
With all these considerations, we obtain the results that are  fully consistent with those from our model calculations,
further supporting the existence of the quadratic SO term.
\fufu{For more details, including contributions from Rashba SO terms,
see Sec.~I of the supplementary material (SM). }
\paragraph*{System and self-consistent SO couplings.---}We consider 2DEGs confined
in \emph{n}-doped ${\rm Al_{0.48}In_{0.52}As/Ga_{0.47}In_{0.53}As}$ \fufu{quantum well} [Fig.~\ref{fig1}(a)], similar to experimental samples of Ref.~\cite{koga:2002}.
\fufu{The well width is 45 nm,} and  a 6-nm wide doped region  in the ${\rm Al_{0.48}In_{0.52}As}$ layer is positioned  15 nm away \fufu{from the well interface}, with the doping density \fufu{$\rho_d = 2.0 \times 10^{18}$ cm$^{-3}$}. The effective electron masses 
are $m^*=0.073m_0$ for Al$_{0.48}$In$_{0.52}$As and $m^* = 0.043 m_0$ for Ga$_{0.47}$In$_{0.53}$As \cite{vurgaftman:2001}, respectively, with $m_0$ being the bare electron mass. To account for the mass discontinuity across different  layers, we solve the Ben Daniel-Duke equation~\cite{bastard:1989,wang:2020}  self-consistently with  the Poisson equation,  beyond using the usual Schr\"odinger equation.
This yields  eigensolutions  used for obtaining the Dresselhaus terms of both intraband [Eq.~\eqref{intrafield}] and  interband [Eq.~\eqref{interfield}] types.  \fufu{At $V_{g} = 0.1$ eV, we obtain $\beta_1=0.16$ meV nm, $\beta_2=0.32$ meV nm, $\Gamma^{(1)}=0.1$ meV nm, and $\Gamma^{(2)}=1.02$ meV nm$^2$} [Figs.~\ref{fig2}(a) and \ref{fig2}(b)], which are used to explore below the spin-related phenomena.

\paragraph*{Anisotropic Fermi contours, swirling textures and spin hybridizations.---}Figure~\ref{fig2}(c) [\ref{fig2}(d)] show spin-resolved dispersions as functions of $k_x$, illustrating the effects of the  linear $\Gamma^{(1)}$ [quadratic  $\Gamma^{(2)}$] interband  SO term.
For the quadratic term $\Gamma^{(2)}$ [Fig.~\ref{fig2}(d)],  a clear band anticrossing appears between the inner two spin branches, as compared to uncoupled (crossed) bands ($\Gamma^{(2)}=0$) [gray lines in Fig.~\ref{fig2}(d)].
This stems from that  $\Gamma^{(2)}$ couples the same spin states across different bands [Fig.~\ref{fig1}(b)].
Notably,  near the anticrossing point, the spin polarization nearly vanishes, as reflected by diminishing size of the markers.
This suppression of  spin polarization  arises from the \emph{hybridization}  of different spin branches.
In sharp contrast,  when only the linear term $\Gamma^{(1)}$ is present [Fig.~\ref{fig2}(c)], we observe that (i) the band dispersions remain  crossed,  resembling the behavior of uncoupled bands [gray lines in Fig.~\ref{fig2}(c)], and that  (ii) no spin  hybridization occurs.
These distinct spin-resolved features have significant implications for spintronic functionalities, e.g., intrinsic spin-Hall effect~\cite{hernandez:2013}  and efficient spin-charge interconversion~\cite{song1:2021,ferreira2023charge}.
Note that the interband SO coupling  has been experimentally verified in Bi-Ag-Au heterostructures with unusual spin textures~\cite{bentmann:2012}.

In Fig.~\ref{fig2}(f), we show the swirling spin texture traced along the constant-energy contours $E(\mathbf{k})=E_{\rm F}$
[green dashed horizontal line in Fig.~\ref{fig2}(d)]
in the $k_x$-$k_y$ plane under the impact of quadratic term  $\Gamma^{(2)}$. 
The arrows along the contours denote the spin vector field $\langle\bm{\sigma}\rangle$. 
Beyond inducing spin hybridization and band anticrossing, we find that  the presence of $\Gamma^{(2)}$ also leads to pronounced anisotropy in the energy disperson. Such anisotropy is expected to manifest in measurable transport properties, e.g., directional dependence in  the charge conductivity tensor.
And, this behavior stands contrast to  the case with only the linear term $\Gamma^{(1)}$, where  both the energy dispersion and the spin texture
remain  isotropic [Fig.~\ref{fig2}(e)].  

\paragraph*{Novel longitudinal Zitterbewegung.---}The dynamics of electron wave packets with SO interaction exhibits an oscillatory (trembling) motion known as
\emph{Zitterbewegung}~\cite{schliemann2005zitterbewegung,schliemann2006zitterbewegung,zawadzki2005zitterbewegung,wen2024trembling,lavor2021zitterbewegung}. With the help of  Ehrenfest's theorem~\cite{simth:1991,lippmann:1965}, we have,
\begin{equation}
  \frac{\partial\langle x\rangle}{\partial t}=\left\langle\frac{\partial \mathcal{H}_{\rm D}}{\partial p_x}\right\rangle,\frac{\partial\langle y\rangle}{\partial t}=\left\langle\frac{\partial \mathcal{H}_{\rm D}}{\partial p_y}\right\rangle,
  \frac{\partial\langle{\bm{\varsigma}}\rangle}{\partial t}=\frac{i}{\hbar}\langle[\mathcal{H}_{\rm D}, \bm{\varsigma}]\rangle,
  \label{eq:xy}
\end{equation}
where $\bm{\varsigma}$ denotes the \emph{pseudospin-spin} tensor that we define to account for  the presence of interband coupling involving both spin ($\sigma$) and subband ($\tau$; pseudospin) degrees of freedom, $\bm{\varsigma}=\sum_{i,j} \varsigma_{ij} {\mathbf{i}}\otimes \mathbf{ j},~~{\rm with}~\varsigma_{ij}=\tau_i\otimes \sigma_j$.
Here the indices $i,j=0,1,2,3$, with $\tau_0=\sigma_0=\mathds{1}$ (identity matrix), and $\{\tau_{1,2,3}\}$, $\{\sigma_{1,2,3}\}$  the standard Pauli matrices acting in the subband and spin subspaces [Eq.~\eqref{fullH}], respectively.
For simplicity, yet without lack of generality, we determine the expectation value of position operator $\mathbf{r(t)}=\{x(t), y(t)\}$
with respect to plane waves (referring to  `large' wave packets). 
Initially, we consider spin-up electrons injected into both of the two bands along the $y$ axis with group velocity $\mathbf{v}_{\rm g}=\hbar k_{0y}/m^*\hat{\mathbf{y}}$, i.e., $\sigma_z^1(0)=\sigma_z^2(0)=1$, and assume $x(0)=y(0)=0$.
For simplifying notations, in what follows we omit the $\langle \cdot\cdot\cdot \rangle$ notation, such that the position variables ($x$, $y$) and the  pseudospin-spin operators ($\varsigma_{ij}$) [Eq.~\eqref{eq:xy}] are to be interpreted as expectation values, i.e., $x\rightarrow \langle x\rangle$, $y\rightarrow \langle y\rangle$, $\varsigma_{ij} \rightarrow \langle \varsigma_{ij}\rangle$. 

In order to fully capture the interband effect, we first ignore the intraband terms [$\beta_1=\beta_2=0$; Eq.~\eqref{intrafield}].
Interestingly, based on  Eqs.~\eqref{H_SO} and \eqref{eq:xy},  we obtain compact expressions for spatial trajectories of electrons arising from  both the linear ($\Gamma^{(1)}$) and quadratic ($\Gamma^{(2)}$) interband SO contributions [Eq.~\eqref{interfield}].
In the purely linear case, we obtain,
\begin{equation}
  x(t)=\frac{4 k_{0y} {\Gamma^{(1)}}^2\left[\cos(\Omega^{(1)}t)-1\right]}{\hbar^2{\Omega^{(1)}}^2},~~y(t)=\frac{\hbar k_{0y}t}{m^*},
  \label{eq:gamma1dy}
\end{equation}
and for the purely quadratic case, we have,
\begin{equation}
  x(t)=0,~~y(t)=\frac{\hbar k_{0y} t}{m^*}+\frac{4 k_{0y} \Delta \Gamma^{(2)}\left[\cos (\Omega^{(2)}t)-1\right]}{\hbar^2{\Omega^{(1)}}^2},
  \label{eq:gamma2dy}
\end{equation}
where we have defined $\Omega^{(1)}=\sqrt{\Delta^2+4 k_{0y}^2 {\Gamma^{(1)}}^2}/{\hbar}$ and $\Omega^{(2)}=\sqrt{\Delta^2+4 k_{0y}^4 {\Gamma^{(2)}}^2}/{\hbar}$, with $\Delta=2\varepsilon_-$ [Eq.~\eqref{fullH}].
From Eq.~\eqref{eq:gamma1dy}, we find that  for the linear term, the \emph{Zitterbewegung} motion is always perpendicular to the initial
group velocity $\mathbf{v}_{\rm g}$ ($\propto {k_{0y}}$),  referring to usual scenario of the so-called transverse \emph{Zitterbewegung} [Fig.~\ref{fig3}(a); Figs.~S3(a) and S3(b) in the SM]. This behavior resembles the \emph{Zitterbewegung} induced by the   \emph{single-band} Dresselhaus SO interaction (see the SM; Sec.~II). Remarkably, for the quadratic term, we uncover an unconventional \emph{Zitterbewegung} that occurs  along the  longitudinal ($k_y$) direction (aligned with $\mathbf{v}_{\rm g}$), while the trajectory  along the transverse ($k_x$) direction remains pinned at zero [Fig.~\ref{fig3}(b)].
This intriguing trembling motion is  attributed to the unique symmetry  of quadratic SO field with $ \tau_{y}\otimes\bm{\sigma}\cdot\mathbf{B}^{\rm D(2)}_{12}(\mathbf{k})$ [Eq.~\eqref{H_SO}].  
\paragraph*{Opposite spin dynamics between different bands.---}
In Figs.~\ref{fig3}(c) and \ref{fig3}(d), we show time evolution of the three spin components $\sigma_{x,y,z}^\nu$ for each band ($\nu=1,2$),  under the influence of  linear and quadratic  interband SO couplings, respectively. 
In the presence  of linear term  $\Gamma^{(1)}$ [Fig.~\ref{fig3}(c)], we observe symmetric (identical) spin evolution in both the $x$ and $z$ components across the two bands, namely $\sigma_x^1=\sigma_x^2$ and $\sigma_z^1=\sigma_z^2$. In contrast, the  $y$ components of the two-band electron spins evolve antisymmetrically with $\sigma_y^1=-\sigma_y^2$.
This behavior reflects the presence of an effective internal magnetic field induced by $\Gamma^{(1)}$, which acts in opposite directions on the two bands
along the  $y$ direction, while preserving identical spin precession dynamics within the $x$-$z$ plane.

In contrast,  under the quadratic term $\Gamma^{(2)}$ [Fig.~\ref{fig3}(d)],  we find that both  $\sigma_x^\nu$ and $\sigma_y^\nu$ remain pinned at their initial value---zero---for both bands.
Remarkably,  the $z$ component of the electron spins for the two bands exhibit opposite time evolution.
Specifically,  $\sigma_z^1$ and $\sigma_z^2$ evolve in opposite trend relative to their initial spin polarization, implying  a spin separation effect between the two bands induced by the quadratic coupling. Since $\sigma_{x,y}^1$ and $\sigma_{x,y}^2$ are locked at zero, the overall spins $\bm{\sigma}^\nu$ of the two bands essentially have opposite dynamics. 

To gain deeper insight into spin dynamics under interband SO coupling,  we decompose the spin evolution by
factoring the \emph{pseudospin-spin} tensor $\varsigma_{ij}$. Specifically, we find that the combinations $(\varsigma_{01}\pm\varsigma_{31})\rightarrow \sigma_x^{1(2)}$, $(\varsigma_{02}\pm\varsigma_{32})\rightarrow \sigma_y^{1(2)}$, and $(\varsigma_{03}\pm\varsigma_{33})\rightarrow \sigma_z^{1(2)}$.
This decomposition enables us to derive  closed-form analytical expressions of the three spin components $\sigma_{x,y,z}^\nu$ of two bands  (see the SM; Sec.~III),  in the presence of either  linear or quadratic interband SO term, justifying our numerical outcomes. 
Also, as expected, we reveal that  the oscillation period of spin is the same as that for the \emph{Zitterbewegung} in real space,
cf. Eq.~(S58) in the SM and Eq.~\eqref{eq:gamma2dy}. 

\paragraph*{Intertwined effect between linear and quadratic  interband terms.---}
When both $\Gamma^{(1)}$ and $\Gamma^{(2)}$ are present, the resulting spin dynamics exhibit novel features that go beyond those arising from either term alone---a phenomenon we refer to as \emph{intertwined} effect of the two interband SO contributions.
 For the in-plane spin components, the contrasting roles of the two types of couplings become particularly evident.  While the quadratic term $\Gamma^{(2)}$ alone yields vanishing in-plane spin components and the linear term $\Gamma^{(1)}$ alone results in antisymmetric $y$ component between the two bands,  we reveal that the presence of  both $\Gamma^{(1)}$ and $\Gamma^{(2)}$ breaks this  antisymmetry in  $\sigma_y^\nu$, as shown in Fig.~\ref{fig4}(a).
This behavior originates from distinct symmetries associated with the linear and quadratic interband SO fields [Eqs.~\eqref{intrafield} and \eqref{interfield}].
For the $x$ component,  although the evolution remains symmetric ($\sigma_x^1=\sigma_x^2$) between the two bands,  the dynamical profile is markedly different from that observed under  $\Gamma^{(1)}$ only, cf. Figs.~\ref{fig4}(a) and \ref{fig3}(c).  

In addition, the out-of-plane $z$ component $\sigma_z^\nu$ further exemplifies the intertwined nature of the dynamics.
When both couplings coexist, we find that the evolution symmetry is broken entirely:  $\sigma_z^1$ and $\sigma_z^2$ evolve in a fashion that is neither symmetric nor antisymmetric [Fig.~\ref{fig4}(a)]. Note that, with only $\Gamma^{(1)}$,  the evolution is symmetric ($\sigma_z^1 = \sigma_z^2$), whereas with  only $\Gamma^{(2)}$,  the dynamics becomes antisymmetric, with $\sigma_z^1$ and  $\sigma_z^2$ evolving in opposite directions from their initial value of 1. 
As a complement, regarding intertwined effect of  $\Gamma^{(1)}$ and $\Gamma^{(2)}$  on spatial \emph{Zitterbewegung}, we find
that their coexistence gives rise to combined trembling motions in both the longitudinal and transverse directions [Fig.~\ref{fig4}(b); Fig.~S4 in the SM].

For completeness, we finally  consider the scenario in which both intraband ($\beta_\nu$, $\nu=1,2$) and interband ($\Gamma^{(1)}$; $\Gamma^{(2)}$) terms are present.
In the case of ``$\Gamma^{(1)}\&~\beta_\nu$'', since both $\Gamma^{(1)}$ and
$\beta_\nu$ are linear in momentum,  we find that the symmetric features of spin dynamics are preserved, with $\sigma_z^1=\sigma_z^2$ and $\sigma_x^1=\sigma_x^2$ [see the SM; Fig.~S6(a)]. In contrast,  in  the case of ``$\Gamma^{(2)}\&~\beta_\nu$'', the opposite charateristics between $\sigma_z^1$ and
$\sigma_z^2$ is broken [see the SM; Fig.~S6(b)], due to  the interplay between the quadratic interband and the  linear intraband couplings. 


\paragraph*{Concluding remarks.---}
We have uncovered a \emph{quadratic} SO contribution that emerges from interband effects in semiconductor heterostructures, \fu{justified by both our model calculations and symmetry analysis along with \gerson{the method of invariants}.}
We have also found that this unusual \emph{even-order} SO term  induces a rich set of spin-related phenomena, including hybridized spin textures, anisotropic dispersions,  avoided band crossings, and novel quantum dynamics in both real space and spin space.
These intriguing  features are not only distinct from those associated with the linear terms, but also highlight new pathways for tailoring spin behavior through interband effects. Our work uncovers a previously overlooked facet of SO  physics and points toward novel strategies for  spintronic functionalities that leverage interband effect and  unusual  SO terms of \emph{even} orders, thus holding broad interest from both theoretical and experimental communities.
\fufu{As a final remark, even in the presence of Rashba-type contributions---both intra- and interband---the essential features induced by the quadratic SO term persist (see the SM; Sec.~IV),  further potentially broadening the frontier of spintronic and even orbitronic functionalities.}
\label{sec:summary}
\paragraph*{Acknowledgments.---} This work was supported by the National Natural Science Foundation of China (Grants No.~12274256, No.~11874236, No.~12022413, No.~11674331, and No.~61674096), the Major Basic Program of Natural Science Foundation of Shandong Province (Grant No.~ZR2021ZD01),  the National Key R\&D Program of China (Grant No.~2022YFA1403200), and  the ``Strategic Priority Research Program (B)'' of the Chinese Academy of Sciences (Grant No.~XDB33030100).


\begin{thebibliography}{38}
\bibitem [{foo({\natexlab{b}})}]{footnote-second-order}%
  \BibitemOpen
  \href@noop {}  \bibinfo {note} {Equation~\eqref{fullH}
  retains SO terms up to the second order in $\mathbf{k}$, in line with our
  focus on capturing the quadratic contribution.}\BibitemShut {Stop}%
\end{thebibliography}
%

\end{document}